\documentclass[onecolumn,useAMS,usenatbib]{mn2e}

\usepackage[dvips,draft]{graphicx}

\title[General relativistic radiation hydrodynamics]
{Equations of general relativistic radiation hydrodynamics 
in Kerr space-time}
\author[Rohta Takahashi]{Rohta Takahashi$^{1,2}$\thanks{E-mail:
rohta@ea.c.u-tokyo.ac.jp}\\
$^{1}$Graduate School of Arts and Sciences, University of Tokyo, Komaba, 
Meguro, Tokyo 153-8902, Japan
\\
$^{2}$Montana State University, 
Department of Physics, Bozeman, MT 59717-0350
}

\begin{document}

\date{Accepted 200X December 15. Received 200X December 14; 
in original form 200X October 11}

\pagerange{\pageref{firstpage}--\pageref{lastpage}} \pubyear{2007}

\maketitle

\label{firstpage}

\begin{abstract}
Equations of fully general relativistic radiation hydrodynamics 
in Kerr space-time are derived. 
While the interactions between matter and radiation are introduced 
in the comoving frame, the derivatives used when describing 
the global evolutions of both the matter and the radiation 
are given in the Boyer-Lindquist frame (BLF) which is a frame 
fixed to the coordinate describing the central black hole. 
Around a rotating black hole, 
both the matter and the radiation are influenced by the frame-dragging 
effects due to the black hole's rotation. 
As a fixed frame, we use the locally non-rotating reference frame (LNRF) 
which is one of the orthonormal frame. 
While the special relativistic effects such as beaming effects 
are introduced by the Lorentz transformation connecting the comoving frame 
and the LNRF, the general relativistic effects such as frame-dragging and 
gravitational redshift are introduced by the tetrads connecting the LNRF 
and the BLF. 
%
%
\end{abstract}


\section{Introduction}
Accretion disks around black holes play one of the essential roles 
in the active universe. 
When the mass accretion rate is near or over the Eddington mass accretion 
rate, the radiation in the disk interact with matter. 
For the supercritical accretion flows considered in e.g. Syfert galaxies 
or the black hole binaries, the photons interact with matter. 
On the other hand, for the hypercritical accretion flows considered 
as one of the leading candidate for 
the central engine of gamma-ray bursts, the neutrinos interact 
with matter around the central black hole. 
In such situations, the dynamics and the energy balance in matter 
and radiation are influenced by each other. 
The radiative transfer in the curved space is investigated 
by many authors 
(e.g. Lindquist 1966; Anderson \& Spiegel 1972; Schmid-Burgk 1978; 
Thorne 1981; Schinder, Bludman \& Piran 1988; Turolla \& Nobili 1988; 
Anile \& Romano 1992; Cardall \& Mezzacappa 2003; Park 2006).  
%
\cite{L66} gives a general treatment of the radiation transfer equation and 
the radiation hydrodynamic equations derived from the zeroth and the first 
moments of the radiation transfer equations by using a comoving Lagrangian 
frame of reference. 
For the higher moments, \cite{AS72} and \cite{T81} gives a detailed 
treatments. 
Especially, \cite{T81} derive general relativistic moment equations 
up to an arbitrary order by introducing projected symmetric 
trace-free (PSTF) tensors. 
This formalism has been used in spherically symmetric problems 
(e.g. Flammang 1982, 1984; Turolla \& Nobili 1988). 
%
Other forms of the moment equations are presented by 
\cite{S88} and \cite{SB89} in a static, spherical space-time 
by using a Lagrangian comoving coordinate system. 
In these formalisms based on the comoving frames, 
the physical quantities in terms of matter and radiation and 
the directional derivatives are described in the comoving frame. 
On the other hand, \cite{M80} introduce the alternative approach  
by using the radiation hydrodynamic equations in the Eulerian framework 
where the derivatives become much simpler form in this framework.  
In this approach, the physical quantities in terms of 
matter and radiation and the derivatives were introduced in the frame 
fixed to the coordinate of the central object, e.g. a black hole, 
while the interactions between matter and radiation were calculated 
in the comoving frame. 
That is,  
the local processes like the interaction between the matter and the 
radiations are evaluated in the comoving frame, while 
the derivatives which are used for the calculations of the global 
dynamics of both the matter and the radiation are derived in the 
frame fixed to the coordinate describing the central object.  
In principle, this formalism can be extended to arbitrary space-time. 
Actually, 
by using the energy-momentum tensor in a covariant form,  
\cite{P93} derived the radiation hydrodynamic equations for 
spherically symmetric systems, and 
recently, \cite{P06} give the explicit 
expressions for the basic equations of the general relativistic 
radiation hydrodynamics in Schwarzschild space-time.   
So far, the formalism in the space-time around a rotating black hole 
which is usually considered for the black holes in the real world 
have not been presented. 
One of the natural next step is to derive the explicit expressions 
for the general relativistic radiation hydrodynamics in space-time 
around a rotating black hole in this formalism. 
In this paper, we assume $c=1$ in most equations except in a few cases 
where $c$ is explicitly used for clarity. 
After giving the general forms of the basic equations for 
the general relativistic radiation hydrodynamics in \S 2, in \S 3 
by using the orthonormal tetrads fixed to the coordinate (\S 3.1),  
the radiation moments (\S 3.2), the radiation four-force (\S 3.3) 
and the radiation hydrodynamic equations (\S 3.4)
in Kerr space-time are derived. 
%
Concluding remarks are given in the last section. 
%

\section{Basic equations for general relativistic radiation hydrodynamic in 
covariant form}
Here, we summarize the basic equations for the general relativistic 
radiation hydrodynamic in covariant form (e.g. Mihalas \& Miharas 1984). 
The energy-momentum tensor for matter of an ideal gas is described as 
%
\begin{equation}
T^{\alpha\beta}=\rho_0 h_{\rm g}u^\alpha u^\beta +P_{\rm g}g^{\alpha\beta},
\end{equation}
%
where $u^\alpha$, $\rho_0$, $h_{\rm g}$ and $P_{\rm g}$ are 
the four-velocity, the rest-mass density, 
the relativistic specific enthalpy and the pressure of the gas, 
respectively. 
The specific enthalpy of the gas, $h_{\rm g}$, is calculated as 
$h_{\rm g}=(\varepsilon_{\rm g}+P_{\rm g})/\rho_0$ 
where $\varepsilon_{\rm g}$ is the energy density of the gas. 
Here, the fluid quantities 
$h_{\rm g}$, $\varepsilon_{\rm g}$, $P_{\rm g}$ and $\rho_0$ are 
all being measured in the comoving frame of the fluid. 
The radiation stress-energy tensor is given as 
%
\begin{equation}
R^{\alpha\beta}=\int\int I({\bf n},\nu)n^\alpha n^\beta d\nu d\Omega ,
\end{equation}
%
where $I(x^\alpha;{\bf n},\nu)$ is the specific intensity of photons 
moving in direction ${\bf n}$ with the frequency $\nu$. 
The photon four-momentum $p^\alpha=(h\nu)(1,{\bf n})$ and 
$n^\alpha\equiv p^\alpha/h\nu$. 
The continuity equation, i.e. particle number conservation equation, 
in the absence of particle creation and annihilation is given 
%
\begin{equation}
(nu^\alpha)_{;\alpha}=0,   
\end{equation}
%
where $n$ is the particle number density measured in the comoving frame. 
Here, $n$ is related to $\rho_0$ as $n=\rho_0/m_{\rm g}$ where 
$m_{\rm g}$ is the mass of the gas particle.  
On the other hand, the conservation equation for the total energy-momentum 
of gas plus radiation is given as 
%
\begin{equation}
(T^{\alpha\beta}+R^{\alpha\beta})_{;\beta}=0. 
\end{equation}
%
The radiation four-force density acting on the matter is given as 
%
\begin{equation}
G^\alpha=\frac{1}{c}\int \int [\chi_\nu I({\bf n},\nu)-\eta_\nu]n^\alpha 
d\nu d\Omega ,
\end{equation}
%
where $\chi_\nu$ and $\eta_\nu$ are the opacity and the emissivity, 
respectively.  
The invariant emissivity and invariant opacity are $\eta_\nu/\nu^2$ and 
$\nu\chi_\nu$, respectively. 
The dynamical equations for the matter and the radiation field are 
described as 
%
\begin{equation}
T^{\alpha\beta}_{~~~;\beta}= G^\alpha,
\end{equation}
%
and
%
\begin{equation}
R^{\alpha\beta}_{~~~;\beta}=-G^\alpha,  
\end{equation}
%
respectively. 
%

\section{Radiation Hydrodynamics in Kerr space-time}
In order to see the correspondence with the work by \cite{P06} 
deriving the equations for the radiation hydrodynamics in the Schwarzschild 
space-time, in the present study we use the Boyer-Lindquist coordinate for 
the description of the rotating black hole's space-time.  
The background geometry around the rotating black hole written by 
the Boyer-Lindquist coordinate is described as 
%
\begin{eqnarray}
ds^2
&=&g_{\alpha\beta}dx^\alpha dx^\beta,\nonumber\\
&=&-\alpha^2 dt^2 +\gamma_{ij}(dx^i+\beta^i dt)(dx^j+\beta^j dt), 
\end{eqnarray}
%
where $i$, $j=r$, $\theta$, $\phi$ and the nonzero components of the lapse 
function $\alpha$, the shift vector $\beta^i$ and the spatial matrix 
$\gamma_{ij}$ are given in the geometric unit as 
\begin{eqnarray}
\alpha=\sqrt{\frac{\Sigma \Delta}{A}},~
\beta^\phi=-\omega,~
\gamma_{rr}=\frac{\Sigma}{\Delta},~
\gamma_{\theta\theta}=\Sigma,~
\gamma_{\phi\phi}=\frac{A\sin^2\theta}{\Sigma}. 
\end{eqnarray}
Here, we use the geometric mass $m=GM/c^2$, 
$\Sigma=r^2+a^2\cos^2\theta$, $\Delta=r^2-2Mr+a^2$ and 
$A=(r^2+a^2)^2-a^2\Delta\sin^2\theta=\Sigma\Delta+2mr(r^2+a^2)$, 
where $M$ is the black hole mass, $G$ is the gravitational constant and 
$c$ is the speed of light. 
The position of the outer and inner horizon, $r_\pm$, 
is calculated from $\Delta=0$ 
as $r_\pm=m\pm(m^2-a^2)^{1/2}$.    
The angular velocity of the frame dragging due to the black hole's 
rotation is calculated as $\omega=-g_{t\phi}/g_{\phi\phi}=2mar/A$. 
%
%
%
%
When $a/m=0$, we obtain 
$\alpha=\Gamma$, $\beta^\phi=0$, $\gamma_{rr}=1/\Gamma^2$, 
$\gamma_{\theta\theta}=r^2$ and $\gamma_{\phi\phi}=r^2\sin^2\theta$ 
where $\Gamma\equiv (1-2m/r)^{1/2}$ \citep{P06}.
%
%
%

\subsection{Orthonormal Tetrads}
In the present study, we use three frames: 
(1) the Boyer-Lindquist frame (BLF) which is the frame 
based on the Boyer-Lindquist coordinate describing the metric and 
have been frequently used for the description of the global dynamics  
(e.g. Gammie \& Popham 1998), 
(2) the locally non-rotating reference frame (LNRF) 
which is the orthonormal frame fixed to the coordinate and 
is called as a fixed frame in \cite{P06},  
and 
(3) the comoving frame where the interactions between the matter 
and the radiation are introduced. 
As a fixed tetrad which is the orthonormal tetrad fixed with respect to 
the coordinates, we use the tetrad for the LNRF \citep{BPT72,FN98}. 
This frame is also called as a zero angular momentum observer (ZAMO) frame. 
This frame corresponds  
the generalization of the fixed frame used in \cite{P06} 
for the space-time around a rotating black hole.  
When using the Boyer-Lindquist coordinate, the LNRF is the frame where 
the fiducial observer in this frame is rotating around the black hole with 
the angular velocity $\omega$ of the frame-dragging due to the black hole's 
rotation with zero radial velocity. 
The base ${\bf e}_{\hat{\alpha}}=\partial/\partial x^{\hat{\alpha}}$ can 
be expressed by the coordinate base $\partial/\partial x^\mu$ as 
%
\begin{equation}
\frac{\partial}{\partial\hat{t}}= 
	\frac{1}{\alpha}\left(\frac{\partial}{\partial t}
	-\beta^\phi\frac{\partial}{\partial \phi}\right),~~~
\frac{\partial}{\partial\hat{r}}= 
	\frac{1}{\sqrt{\mathstrut \gamma_{rr}}}
	\frac{\partial}{\partial r},~~~
\frac{\partial}{\partial\hat{\theta}}= 
	\frac{1}{\sqrt{\mathstrut \gamma_{\theta\theta}}}
	\frac{\partial}{\partial \theta},~~~
\frac{\partial}{\partial\hat{\phi}}= 
	\frac{1}{\sqrt{\mathstrut \gamma_{\phi\phi}}}
	\frac{\partial}{\partial \phi}. 
\end{equation}
%
Here, $(x^{\hat{0}},~x^{\hat{1}},~x^{\hat{2}},~x^{\hat{3}})
=(\hat{t},~\hat{r},~\hat{\theta},~\hat{\phi})$. 
In the LNRF, 
a fiducial observer who is fixed with respect to the coordinates 
see the fluid with the three velocity whose components are 
calculated as 
%
\begin{equation}
\hat{v}^i=\frac{u^{\hat{i}}}{u^{\hat{t}}},~~~(i=r,~\theta,~\phi)
\end{equation}
%
where $u^{\hat{\alpha}}$ is the four-velocity in the LNRF. 
The hat denotes the physical quantities measured in the LNRF. 
The components of the three velocity are explicitly 
calculated as 
%
\begin{equation}
\hat{v}^r=\frac{\sqrt{\mathstrut \gamma_{rr}}}{\alpha u^t}u^r,~~~
\hat{v}^\theta
	=\frac{\sqrt{\mathstrut \gamma_{\theta\theta}}}{\alpha u^t}u^\theta,~~~
\hat{v}^r=\frac{\sqrt{\mathstrut \gamma_{\phi\phi}}}{\alpha}
	(\Omega+\beta^\phi),~~~
\end{equation}
%
where $\Omega\equiv u^\phi/u^t$ is the angular velocity and 
we have used $u^{\hat{t}}=-u_{\hat{t}}=\alpha u^t$. 
The Lorentz factor $\hat{\gamma}$ for this three velocity is 
calculated as 
%
\begin{equation}
\hat{\gamma}\equiv(1-\hat{v}^2)^{-1/2}=\alpha u^t, 
\end{equation}
%
where $\hat{v}^2={\bf v}\cdot{\bf v}=\hat{v}_i\hat{v}^i=\hat{v}_r^2+\hat{v}_\theta^2+\hat{v}_\phi^2$. 

A tetrad base for the comoving frame 
$\partial/\partial x^{\bar{\alpha}}$
is calculated by 
the Lorentz transformation as 
%
\begin{equation}
\frac{\partial}{\partial x^{\bar{\alpha}}}
=\Lambda^{\hat{\beta}}_{~\bar{\alpha}}({\bf v})
\frac{\partial}{\partial x^{\hat{\beta}}}, 
\end{equation}
%
where the bar denotes the physical quantities measured 
in the comoving frame. 
The components of the Lorentz transformation 
$\Lambda^{\hat{\alpha}}_{~\bar{\beta}}({\bf v})$ are given as 
$\Lambda^{\hat{t}}_{~\bar{t}}=\hat{\gamma}$, 
$\Lambda^{\hat{i}}_{~\bar{t}}=\hat{\gamma} \hat{v}^i$, 
$\Lambda^{\hat{t}}_{~\bar{j}}=\hat{\gamma} \hat{v}_j$ and 
$\Lambda^{\hat{i}}_{~\bar{j}}=\delta^i_{~j}+\hat{v}^i \hat{v}_j \hat{\gamma}^2/(1+\hat{\gamma})$ 
($i,~j=r,~\theta,~\phi$).   
Here, $(x^{\bar{0}},~x^{\bar{1}},~x^{\bar{2}},~x^{\bar{3}})
=(\bar{t},~\bar{r},~\bar{\theta},~\bar{\phi})$. 
The components of the base of the comoving tetrad 
$\partial/\partial x^{\bar{\alpha}}$
can be expressed by the coordinate base 
$\partial/\partial x^{\alpha}$
as 
%
\begin{eqnarray}
\frac{\partial}{\partial\bar{t}}&=&
	\frac{\hat{\gamma}}{\alpha} \frac{\partial}{\partial t}
	+\frac{\hat{\gamma} \hat{v}_r}{\sqrt{\mathstrut \gamma_{rr}}}
		\frac{\partial}{\partial r}
	+\frac{\hat{\gamma} \hat{v}_\theta}
			{\sqrt{\mathstrut \gamma_{\theta\theta}}}
		\frac{\partial}{\partial \theta}
	+\hat{\gamma}\left(
		\frac{ \hat{v}_\phi}
			{\sqrt{\mathstrut \gamma_{\phi\phi}}}
		-\frac{\beta^\phi}{\alpha}
	\right)
		\frac{\partial}{\partial \phi},\nonumber\\
\frac{\partial}{\partial\bar{r}}&=&
	\frac{\hat{\gamma} \hat{v}_r}{\alpha} \frac{\partial}{\partial t}
	+\frac{1}{\sqrt{\mathstrut \gamma_{rr}}}
		\left[1+\hat{v}_r^2\left(
				\frac{\hat{\gamma}^2}{\hat{\gamma}+1}\right)\right] 
		\frac{\partial}{\partial r}
	+\frac{\hat{v}_r \hat{v}_\theta}
			{\sqrt{\mathstrut \gamma_{\theta\theta}}}
		\left(\frac{\hat{\gamma}^2}{\hat{\gamma}+1}\right)
		\frac{\partial}{\partial \theta}
	+\left[
		\frac{\hat{v}_r \hat{v}_\phi}{\sqrt{\mathstrut \gamma_{\phi\phi}}}
		\left(\frac{\hat{\gamma}^2}{\hat{\gamma}+1}\right)
		-\frac{\beta^\phi}{\alpha}\hat{\gamma} \hat{v}_r
	\right]
		\frac{\partial}{\partial \phi},\nonumber\\
\frac{\partial}{\partial\bar{\theta}}&=&
	\frac{\hat{\gamma} \hat{v}_\theta}{\alpha} \frac{\partial}{\partial t}
	+\frac{\hat{v}_r \hat{v}_\theta}{\sqrt{\mathstrut \gamma_{rr}}}
		\left(\frac{\hat{\gamma}^2}{\hat{\gamma}+1}\right)
		\frac{\partial}{\partial r}
	+\frac{1}{\sqrt{\mathstrut \gamma_{\theta\theta}}}
		\left[1+\hat{v}_\theta^2\left(
			\frac{\hat{\gamma}^2}{\hat{\gamma}+1}\right)\right] 
		\frac{\partial}{\partial \theta}
	+\left[
		\frac{\hat{v}_\theta \hat{v}_\phi}
				{\sqrt{\mathstrut \gamma_{\phi\phi}}}
		\left(\frac{\hat{\gamma}^2}{\hat{\gamma}+1}\right)
		-\frac{\beta^\phi}{\alpha}\hat{\gamma} \hat{v}_\theta
	\right]
		\frac{\partial}{\partial \phi},\nonumber\\
\frac{\partial}{\partial\bar{\phi}}&=&
	\frac{\hat{\gamma} \hat{v}_\phi}{\alpha} \frac{\partial}{\partial t}
	+\frac{\hat{v}_r \hat{v}_\phi}{\sqrt{\mathstrut \gamma_{rr}}} 
		\left(\frac{\hat{\gamma}^2}{\hat{\gamma}+1}\right)
		\frac{\partial}{\partial r}
	+\frac{\hat{v}_\phi \hat{v}_\theta}
			{\sqrt{\mathstrut \gamma_{\theta\theta}}}
		\left(\frac{\hat{\gamma}^2}{\hat{\gamma}+1}\right)
		\frac{\partial}{\partial \theta}
	+\left\{
		\frac{1}{\sqrt{\mathstrut \gamma_{\phi\phi}}}
		\left[1+
			\hat{v}_\phi^2 \left(\frac{\hat{\gamma}^2}{\hat{\gamma}+1}\right)
		\right]
		-\frac{\beta^\phi}{\alpha}\hat{\gamma} \hat{v}_\phi
	\right\}
		\frac{\partial}{\partial \phi}.
\end{eqnarray}
%
In the same way, the inverse transformation from the base of 
the comoving tetrad to the coordinate base is calculated by using 
the inverse Lorentz transformation 
$\Lambda^{\bar{\alpha}}_{~\hat{\beta}}(-{\bf v})$, and 
the coordinate base $\partial/\partial x^{\alpha}$ 
is calculated from the base of the comoving tetrad 
$\partial/\partial x^{\bar{\alpha}}$ as 
%
\begin{eqnarray}
\frac{1}{\alpha}
\frac{\partial}{\partial t}&=&
	\left(
		\hat{\gamma}-\frac{\beta^\phi\sqrt{\mathstrut \gamma_{\phi\phi}}}
			{\alpha}\hat{\gamma} \hat{v}_\phi
	\right)
		\frac{\partial}{\partial \bar{t}}
	+\left[
		-\hat{\gamma} \hat{v}_r
		+\frac{\beta^\phi\sqrt{\mathstrut \gamma_{\phi\phi}}}{\alpha}
			\hat{v}_r \hat{v}_\phi
			\left(\frac{\hat{\gamma}^2}{\hat{\gamma}+1}\right)
	\right]
		\frac{\partial}{\partial \bar{r}}
	+\left[
		-\hat{\gamma} \hat{v}_\theta
		+\frac{\beta^\phi\sqrt{\mathstrut \gamma_{\phi\phi}}}{\alpha}
		\hat{v}_\theta \hat{v}_\phi
			\left(\frac{\hat{\gamma}^2}{\hat{\gamma}+1}\right)
	\right]
		\frac{\partial}{\partial \bar{\theta}}
	\nonumber\\
	&&
	+\left\{
		-\hat{\gamma} \hat{v}_\phi
		+\frac{\beta^\phi\sqrt{\mathstrut \gamma_{\phi\phi}}}{\alpha}
			\left[
				1+\hat{v}_\phi^2\left(
				\frac{\hat{\gamma}^2}{\hat{\gamma}+1}\right)
			\right]
	\right\}
		\frac{\partial}{\partial \bar{\phi}},\nonumber\\
\frac{1}{\sqrt{\mathstrut \gamma_{rr}}}
\frac{\partial}{\partial r}&=&
	-\hat{\gamma} \hat{v}_r \frac{\partial}{\partial \bar{t}}
	+\left[
		1+\hat{v}_r^2\left(\frac{\hat{\gamma}^2}{\hat{\gamma}+1}\right)
	\right]
		\frac{\partial}{\partial \bar{r}}
	+\hat{v}_r \hat{v}_\theta\left(
			\frac{\hat{\gamma}^2}{\hat{\gamma}+1}\right)
		\frac{\partial}{\partial \bar{\theta}}
	+\hat{v}_r \hat{v}_\phi \left(
						\frac{\hat{\gamma}^2}{\hat{\gamma}+1}\right)
		\frac{\partial}{\partial \bar{\phi}},\nonumber\\
\frac{1}{\sqrt{\mathstrut \gamma_{\theta\theta}}}
\frac{\partial}{\partial \theta}&=&
	-\hat{\gamma} \hat{v}_\theta \frac{\partial}{\partial \bar{t}}
	+\hat{v}_r \hat{v}_\theta\left(
						\frac{\hat{\gamma}^2}{\hat{\gamma}+1}\right)
		\frac{\partial}{\partial \bar{r}}
	+\left[
		1+\hat{v}_\theta^2\left(\frac{\hat{\gamma}^2}{\hat{\gamma}+1}\right)
	\right]
		\frac{\partial}{\partial \bar{\theta}}
	+\hat{v}_\theta \hat{v}_\phi \left(
							\frac{\hat{\gamma}^2}{\hat{\gamma}+1}\right)
		\frac{\partial}{\partial \bar{\phi}},\nonumber\\
\frac{1}{\sqrt{\mathstrut \gamma_{\phi\phi}}}
\frac{\partial}{\partial \phi}&=&
	-\hat{\gamma} \hat{v}_\phi \frac{\partial}{\partial \bar{t}}
	+\hat{v}_r \hat{v}_\phi 
		\left(\frac{\hat{\gamma}^2}{\hat{\gamma}+1}\right)
		\frac{\partial}{\partial \bar{r}}
	+\hat{v}_\theta \hat{v}_\phi\left(\frac{\hat{\gamma}^2}{\hat{\gamma}+1}\right)
		\frac{\partial}{\partial \bar{\theta}}
	+\left[
		1+\hat{v}_\phi^2\left(\frac{\hat{\gamma}^2}{\hat{\gamma}+1}\right)
	\right]
		\frac{\partial}{\partial \bar{\phi}}. 
\end{eqnarray}
%

\subsection{Radiation moments}
The radiation energy density $E$, the radiation flux $F^i$ and 
the radiation pressure tensor $P^{ij}$ are defined as the zeroth, 
the first and the second moments of the specific intensity 
$I_\nu(x^\mu,~{\bf n})$, respectively. 
We denote the radiation moments as 
%
%
%
\begin{equation}
     \hat{E}=\int\int \hat{I}_{\hat{\nu}} d\hat{\nu} d\hat{\Omega},~~~
 \hat{F}^{i}=\int\int \hat{I}_{\hat{\nu}} 
				\hat{n}^i d\hat{\nu} d\hat{\Omega},~~~
\hat{P}^{ij}=\int\int \hat{I}_{\hat{\nu}} 
				\hat{n}^i \hat{n}^j d\hat{\nu} d\hat{\Omega},
\end{equation}
when measured in the LNRF, 
and  
%
\begin{equation}
     \bar{E}=\int\int \bar{I}_{\bar{\nu}} d\bar{\nu} d\bar{\Omega},~~~
 \bar{F}^{i}=\int\int \bar{I}_{\bar{\nu}} 
				\bar{n}^i d\bar{\nu} d\bar{\Omega},~~~
\bar{P}^{ij}=\int\int \bar{I}_{\bar{\nu}} 
				\bar{n}^i \bar{n}^j d\bar{\nu} d\bar{\Omega}, 
\end{equation}
when measured in the comoving frame. 
Correspondingly, the radiation stress tensors 
for 
the LNRF and the comoving frame are given as 
%
\begin{equation}
R^{\hat{\alpha}\hat{\beta}}=
	\left(
	\begin{array}{cccc}
	\hat{E} & \hat{F}^r 
		& \hat{F}^\theta & \hat{F}^\phi \\
	\hat{F}^r & \hat{P}^{rr} 
		& \hat{P}^{r\theta} & \hat{P}^{r\phi} \\
	\hat{F}^\theta & \hat{P}^{r\theta} 
		& \hat{P}^{\theta\theta} & \hat{P}^{\theta\phi} \\
	\hat{F}^\phi & \hat{P}^{r\phi} 
		& \hat{P}^{\theta\phi} & \hat{P}^{\phi\phi} 	
	\end{array}
	\right),
\end{equation}
%
and
%
\begin{equation}
R^{\bar{\alpha}\bar{\beta}}=
	\left(
	\begin{array}{cccc}
	\bar{E} & \bar{F}^r 
		& \bar{F}^\theta & \bar{F}^\phi \\
	\bar{F}^r & \bar{P}^{rr} 
		& \bar{P}^{r\theta} & \bar{P}^{r\phi} \\
	\bar{F}^\theta & \bar{P}^{r\theta} 
		& \bar{P}^{\theta\theta} & \bar{P}^{\theta\phi} \\
	\bar{F}^\phi & \bar{P}^{r\phi} 
		& \bar{P}^{\theta\phi} & \bar{P}^{\phi\phi} 	
	\end{array}
	\right),
\end{equation}
%
respectively. 
The contravariant components of the radiation stress tensor 
$R^{\alpha\beta}$ are calculated from $R^{\hat{\alpha}\hat{\beta}}$ 
by the transformation as 
%
\begin{equation}
R^{\alpha\beta}=
	\frac{\partial x^\alpha}{\partial x^{\hat{\mu}}}
	\frac{\partial x^\beta}{\partial x^{\hat{\nu}}}
	R^{\hat{\mu}\hat{\nu}},  
\end{equation}
%
and explicitly given as 
%
\begin{eqnarray}
R^{\alpha\beta}=
\left[
	\begin{array}{cccc}
	\displaystyle 
	\frac{\hat{E}}{\alpha^2} 
		& \displaystyle 
			\frac{\hat{F}^r}{\alpha\sqrt{\mathstrut \gamma_{rr}}} 
		& \displaystyle 
			\frac{\hat{F}^\theta}
					{\alpha\sqrt{\mathstrut \gamma_{\theta\theta}}}
		& \displaystyle 
			\frac{1}{\alpha}
			\left(
				\frac{\hat{F}^\phi}{\sqrt{\mathstrut \gamma_{\phi\phi}}}
				-\beta^\phi
					\frac{\hat{E}}{\alpha}
			\right)
		\vspace{.5em}
		\\
	\displaystyle 
	\frac{\hat{F}^r}{\alpha\sqrt{\mathstrut \gamma_{rr}}}
		& \displaystyle 
			\frac{\hat{P}^{rr}}{\gamma_{rr}}
		& \displaystyle 
			\frac{\hat{P}^{r\theta}}
				{\sqrt{\mathstrut \gamma_{rr}\gamma_{\theta\theta}}}
		& \displaystyle 
			\frac{1}{\sqrt{\mathstrut \gamma_{rr}}}
			\left(
				\frac{\hat{P}^{r\phi}}{\sqrt{\mathstrut \gamma_{\phi\phi}}}
				-\beta^\phi
					\frac{\hat{F}^r}{\alpha}
			\right)
		\vspace{.5em}
		\\
	\displaystyle 
	\frac{\hat{F}^\theta}{\alpha\sqrt{\mathstrut \gamma_{\theta\theta}}}
		& \displaystyle 
			\frac{\hat{P}^{r\theta}}
				{\sqrt{\mathstrut \gamma_{rr}\gamma_{\theta\theta}}}
		& \displaystyle 
			\frac{\hat{P}^{\theta\theta}}{\gamma_{\theta\theta}}
		& \displaystyle 
			\frac{1}{\sqrt{\mathstrut \gamma_{\theta\theta}}}
			\left(
				\frac{\hat{P}^{\theta\phi}}
						{\sqrt{\mathstrut \gamma_{\phi\phi}}}
				-\beta^\phi\frac{\hat{F}^\theta}{\alpha}
			\right)
		\vspace{.5em}
		\\
	\displaystyle 
			\frac{1}{\alpha}
			\left(
				\frac{\hat{F}^\phi}{\sqrt{\mathstrut \gamma_{\phi\phi}}}
				-\beta^\phi
					\frac{\hat{E}}{\alpha}
			\right)
		& \displaystyle 
			\frac{1}{\sqrt{\mathstrut \gamma_{rr}}}
			\left(
				\frac{\hat{P}^{r\phi}}{\sqrt{\mathstrut \gamma_{\phi\phi}}}
				-\beta^\phi
					\frac{\hat{F}^r}{\alpha}
			\right)
		& \displaystyle 
			\frac{1}{\sqrt{\mathstrut \gamma_{\theta\theta}}}
			\left(
				\frac{\hat{P}^{\theta\phi}}
						{\sqrt{\mathstrut \gamma_{\phi\phi}}}
				-\beta^\phi\frac{\hat{F}^\theta}{\alpha}
			\right)
		& \displaystyle 
			\frac{\hat{P}^{\phi\phi}}{\gamma_{\phi\phi}}
			-\frac{2\beta^\phi}{\alpha\sqrt{\mathstrut \gamma_{\phi\phi}}}
				\hat{F}^\phi
			+\left(\frac{\beta^\phi}{\alpha}\right)^2 \hat{E}
\end{array}
\right]. \nonumber\\
\label{eq:RmomCFbyLNRF}
\end{eqnarray}
%
By using the Lorentz transformations, 
the radiation moments measured in the comoving frame 
are calculated from those measured in the LNRF as 
\begin{equation}
R^{\bar{\alpha}\bar{\beta}}=
	\Lambda^{\bar{\alpha}}_{~\hat{\mu}}(-{\rm v})
	\Lambda^{\bar{\beta}}_{~\hat{\nu}}(-{\rm v})
	R^{\hat{\mu}\hat{\nu}},  
\end{equation}
and explicitly given as \citep{MM84,P06}
%
\begin{eqnarray}
\bar{E} &=& \hat{\gamma}^2 
		\left(
			\hat{E}-2\hat{v}_i \hat{F}^i +\hat{v}_i \hat{v}_j \hat{P}^{ij} 
		\right),\\
\bar{F}^i &=& -\hat{\gamma}^2 \hat{v}^i \hat{E} 
		+\hat{\gamma}\left[
			\delta^i_j+\left(\hat{\gamma}+\frac{\hat{\gamma}^2}
			{\hat{\gamma}+1}\right)\hat{v}^i \hat{v}_j
		\right]\hat{F}^j
		-\hat{\gamma} \hat{v}_j \left(
			\delta^i_k+\frac{\hat{\gamma}^2}{\hat{\gamma}+1}
			\hat{v}^i \hat{v}_k
		\right) \hat{P}^{jk},\\
\bar{P}^{ij} &=& \hat{\gamma}^2 \hat{v}^i \hat{v}^j \hat{E}
		-\hat{\gamma} \left(
			\hat{v}^i \delta^j_k +\hat{v}^j \delta^i_k 
			+2\frac{\hat{\gamma}^2}{\hat{\gamma}+1} 
			\hat{v}^i \hat{v}^j \hat{v}_k
		\right)\hat{F}^k
		+\left(
			\delta^i_k+\frac{\hat{\gamma}^2}{\hat{\gamma}+1}
			\hat{v}^i \hat{v}_k
		\right)
		\left(
			\delta^j_l+\frac{\hat{\gamma}^2}{\hat{\gamma}+1}
			\hat{v}^j \hat{v}_l
		\right)
		\hat{P}^{kl}. 
\end{eqnarray}
%
Inversely, the radiation moments measured in the LNRF 
are calculated from those measured in the comoving frame as 
\begin{equation}
R^{\hat{\alpha}\hat{\beta}}=
	\Lambda^{\hat{\alpha}}_{~\bar{\mu}}({\rm v})
	\Lambda^{\hat{\beta}}_{~\bar{\nu}}({\rm v})
	R^{\bar{\mu}\bar{\nu}},  
\end{equation}
and explicitly given as 
%
\begin{eqnarray}
\hat{E} &=& \hat{\gamma}^2 
		\left(
			\bar{E}+2\hat{v}_i \bar{F}^i +\hat{v}_i \hat{v}_j \bar{P}^{ij} 
		\right),\\
\hat{F}^i &=& \hat{\gamma}^2 \hat{v}^i \bar{E} 
		+\hat{\gamma}\left[
			\delta^i_j+\left(\hat{\gamma}
			+\frac{\hat{\gamma}^2}{\hat{\gamma}+1}\right)\hat{v}^i \hat{v}_j
		\right]\bar{F}^j
		+\hat{\gamma} \hat{v}_j \left(
			\delta^i_k+\frac{\hat{\gamma}^2}{\hat{\gamma}+1}
			\hat{v}^i \hat{v}_k
		\right) \bar{P}^{jk},\\
\hat{P}^{ij} &=& \hat{\gamma}^2 \hat{v}^i \hat{v}^j \bar{E}
		+\hat{\gamma} \left(
			\hat{v}^i \delta^j_k +\hat{v}^j \delta^i_k 
			+2\frac{\hat{\gamma}^2}{\hat{\gamma}+1} 
			\hat{v}^i \hat{v}^j \hat{v}_k
		\right)\bar{F}^k
		+\left(
			\delta^i_k+\frac{\hat{\gamma}^2}{\hat{\gamma}+1}
			\hat{v}^i \hat{v}_k
		\right)
		\left(
			\delta^j_l+\frac{\hat{\gamma}^2}{\hat{\gamma}+1}
			\hat{v}^j \hat{v}_l
		\right)
		\bar{P}^{kl}. 
\end{eqnarray}
%

\subsection{Radiation four-force density}
The radiation four-force density measured in the comoving frame is 
given as \citep{MM84}
%
\begin{equation}
G^{\bar{\alpha}}=\frac{1}{c}\int \int~
	(\bar{\chi}\bar{I}_{\bar{\nu}}-\bar{\eta})n^{\bar{\alpha}}~
	d\bar{\nu}d\bar{\Omega}, 
\end{equation}
%
where $\bar{\chi}$ and $\bar{\eta}$ are the mean opacity and the emissivity 
measured in the comoving frame, respectively. 
The time component $G^{\bar{t}}$
has the dimension $c^{-1}$ times the net rate of the 
radiation energy per unit volume, and 
the spatial component $G^{\bar{i}}$ has the dimension of the net rate 
of the momentum exchange between the matter and the radiation. 
The time component of the radiation four-force density 
measured in the comoving frame can be calculated as \citep{MM84,P06}
%
\begin{equation}
G^{\bar{t}}=\bar{\Gamma}-\bar{\Lambda}
\end{equation}
%
where the heating function $\bar{\Gamma}$ and the cooling function 
$\bar{\Lambda}$ are defined as 
%
\begin{equation}
\bar{\Gamma}\equiv\frac{1}{c}\int\int~ \bar{\chi}\bar{I}_{\bar{\nu}} 
			d\bar{\nu}d\bar{\Omega},~~~~~
\bar{\Lambda}\equiv\frac{1}{c}\int\int~ \bar{\eta} 
			d\bar{\nu}d\bar{\Omega}.
\end{equation}
%
%
%
%
%
The components of the radiation force $G^\alpha$ is calculated from 
those in the comoving frame $G^{\bar{\alpha}}$ by the transformation 
%
\begin{equation}
G^\alpha=\frac{\partial x^\alpha}{\partial x^{\bar{\beta}}}G^{\bar{\beta}}, 
\end{equation}
%
and explicitly given as 
%
\begin{eqnarray}
G^t &=& \frac{\hat{\gamma}}{\alpha}
		\left(
			G^{\bar{t}}+\hat{v}_i G^{\bar{i}}
		\right),\nonumber\\
G^r &=& \frac{1}{\sqrt{\mathstrut \gamma_{rr}}}
		\left(
			G^{\bar{r}}+\hat{\gamma} \hat{v}_r G^{\bar{t}}
			+\frac{\hat{\gamma}^2}{\hat{\gamma}+1}
			\hat{v}_r \hat{v}_i G^{\bar{i}} 
		\right),\nonumber\\
G^\theta &=& \frac{1}{\sqrt{\mathstrut \gamma_{\theta\theta}}}
		\left(
			G^{\bar{\theta}}+\hat{\gamma} \hat{v}_\theta G^{\bar{t}}
			+\frac{\hat{\gamma}^2}
				{\hat{\gamma}+1}\hat{v}_\theta \hat{v}_i G^{\bar{i}} 
		\right),\nonumber\\
G^\phi &=& \frac{G^{\bar{\phi}}}{\sqrt{\mathstrut \gamma_{\phi\phi}}}
	+\hat{\gamma}\left(
		\frac{\hat{v}_\phi}{\sqrt{\mathstrut \gamma_{\phi\phi}}}
		-\frac{\beta^\phi}{\alpha}
	\right)G^{\bar{t}}
	+\hat{\gamma}\left[
		\frac{\hat{\gamma}\hat{v}_\phi}
			{\sqrt{\mathstrut \gamma_{\phi\phi}}(\hat{\gamma}+1)}
		-\frac{\beta^\phi}{\alpha}
	\right]\hat{v}_i G^{\bar{i}}. 
\end{eqnarray}
%

\subsection{Radiation hydrodynamic equations}
%

\subsubsection{Continuity equation}
As a continuity equation, now we consider the particle number conservation, 
$(nu^\alpha)_{;\alpha}=0$. 
This equation can be calculated as 
%
\begin{equation}
\frac{\partial}{\partial t}\left( n\alpha u^t 
					\sqrt{\mathstrut \gamma}\right) 
+\frac{\partial}{\partial x^i} \left(n\alpha u^i 
					\sqrt{\mathstrut \gamma}\right)=0, 
\end{equation}
%
where 
$\gamma\equiv {\rm det}\gamma_{ij}$. 
In the case of the Kerr metric written by the Boyer-Lindquist coordinate
$\gamma=(\Sigma A/\Delta)\sin^2\theta$. 
%

\subsubsection{Hydrodynamic equations}
The relativistic Euler equations are obtained by the projection of the 
equation of the energy-momentum tensor  
$T^{\alpha\beta}_{~~;\beta}=G^\alpha$ 
on the specific directions by using the projection tensor 
$P^{\alpha\beta}=g^{\alpha\beta}+u^\alpha u^\beta$ as 
$P^\alpha_\beta T^{\beta\delta}_{~~;\delta}=P^\alpha_\beta G^\beta$.    
From this, we can obtain 
$\rho_0 h_{\rm g} u^\alpha_{~;\beta}u^\beta
+g^{\alpha\beta}P_{g,\beta}+u^\alpha u^\beta P_{g,\beta}
=G^\alpha+u^\alpha u_\beta G^\beta$. 
The radial part of the Euler equation, i.e. momentum conservation for 
$r$-direction, is calculated as 
%
\begin{eqnarray}
&&
 \rho_0 h_{\rm g} u^t \frac{\partial u^r}{\partial t}
+\rho_0 h_{\rm g} u^i \frac{\partial u^r}{\partial x^i}
+\frac{\rho_0 h_{\rm g}}{2} 
	\left[
		\left(\ln\gamma_{rr}\right)_{,r}(u^r)^2 
		+2 \left(\ln\gamma_{rr}\right)_{,\theta} u^r u^\theta 
		-\left(\ln\gamma_{\theta\theta}\right)_{,r}
			\frac{\gamma_{\theta\theta}}{\gamma_{rr}}(u^\theta)^2
	\right]
\nonumber\\
&&
-\frac{\rho_0 h_{\rm g}}{2} 
	\frac{(u^t)^2}{\gamma_{rr}} 
	\left[
		\gamma_{\phi\phi}\left(\ln\gamma_{\phi\phi}\right)_{,r}
			(\Omega+\beta^\phi)^2
		+2\gamma_{\phi\phi}\beta^\phi_{,r}(\Omega+\beta^\phi)
		-2\alpha^2\left(\ln\alpha\right)_{,r}
	\right]
+\frac{1}{\gamma_{rr}}\frac{\partial P_g}{\partial r}
+u^r\left(
		u^t \frac{\partial P_g}{\partial t}
		+u^i \frac{\partial P_g}{\partial x^i}
	\right)
\nonumber\\
&&
=-\alpha^2 u^t u^r G^t 
+\left[1+\gamma_{rr}(u^r)^2\right]G^r
+\gamma_{\theta\theta}u^r u^\theta G^\theta
+\gamma_{\phi\phi}(\Omega+\beta^\phi)(G^\phi + \beta^\phi G^t)u^tu^r, 
\label{eq:rEuler}
\end{eqnarray}
%
where the derivatives for $\ln \alpha$, $\beta^\phi$, $\ln \gamma_{rr}$, 
$\ln \gamma_{\theta\theta}$ and $\ln \gamma_{\phi\phi}$ 
with respect to $r$ and $\theta$ are given in App. \ref{app:metric}.  
The terms including $\beta^\phi(=-\omega)$ explicitly show  
the effects of the frame-dragging due to the black hole's rotation. 
Here, we have used the relation 
$g_{tt,r}(u^t)^2+2g_{t\phi,r}u^t u^\phi
+g_{\phi\phi,r}(u^\phi)^2=
	(u^t)^2
	\left[
		\gamma_{\phi\phi}\left(\ln\gamma_{\phi\phi}\right)_{,r}
			(\Omega+\beta^\phi)^2
		+2\gamma_{\phi\phi}\beta^\phi_{,r}(\Omega+\beta^\phi)
		-2\alpha^2\left(\ln\alpha\right)_{,r}
	\right]
$. 
This terms are related to the centrifugal force. 
Actually, this relation can be also calculated as 
$g_{tt,r}(u^t)^2+2g_{t\phi,r}u^t u^\phi
+g_{\phi\phi,r}(u^\phi)^2=
(\ln\gamma_{\phi\phi})_{,r}\gamma_{\phi\phi}(u^t)^2
(\Omega-\Omega_-)(\Omega-\Omega_+)$ where  
$\Omega_\pm$ is calculated as   
%
\begin{eqnarray}
\Omega_\pm 
	&=& -\frac{g_{t\phi,r}}{g_{\phi\phi,r}}
		\pm\sqrt{\left( \frac{g_{t\phi ,r}}{g_{\phi\phi ,r}} \right)^2
				-\frac{g_{tt,r}}{g_{\phi\phi,r}}}
\nonumber\\
	&=&\frac{\pm \sqrt{m}/\sin\theta}
			{\displaystyle \sqrt{\frac{r\Sigma^2}{2r^2-\Sigma}
				-m(r^2+a^2-\Sigma)+ma^2 \sin^2\theta}
				\pm \sqrt{m}a\sin^2\theta}. 
\end{eqnarray}
%
These angular velocity $\Omega_\pm$ is the generalization of the 
Keplerian angular velocity 
$\Omega_{\rm K}^\pm=\pm m^{1/2}/(r^{3/2}\pm a m^{1/2})$, and 
in the limit of $\theta=\pi/2$, 
we obtain $\Omega_\pm = \Omega_{\rm K}^\pm$. 
%
%
In the right hand side of Eq. (\ref{eq:rEuler}), 
the terms not including the derivatives represents 
the gravitational acceleration and 
the centrifugal acceleration, respectively.
The Euler equation in $\theta$-direction, i.e. momentum conservation 
in $\theta$-direction is calculated as 
%
\begin{eqnarray}
&&
 \rho_0 h_{\rm g} u^t \frac{\partial u^\theta}{\partial t}
+\rho_0 h_{\rm g} u^i \frac{\partial u^\theta}{\partial x^i}
+\frac{\rho_0 h_{\rm g}}{2} 
	\left[
		\left(\ln\gamma_{\theta\theta}\right)_{,\theta}(u^\theta)^2
		+2\left(\ln\gamma_{\theta\theta}\right)_{,r}u^r u^\theta 
		-\left(\ln\gamma_{rr}\right)_{,r}
			\frac{\gamma_{rr}}{\gamma_{\theta\theta}}(u^r)^2
	\right]
\nonumber\\
&&
-\frac{\rho_0 h_{\rm g}}{2}
	\frac{(u^t)^2}{\gamma_{\theta\theta}} 
	\left[
		\gamma_{\phi\phi}\left(\ln\gamma_{\phi\phi}\right)_{,\theta}
			(\Omega+\beta^\phi)^2
		+2\gamma_{\phi\phi}\beta^\phi_{,\theta}(\Omega+\beta^\phi)
		-2\alpha^2\left(\ln\alpha\right)_{,\theta}
	\right]
+\frac{1}{\gamma_{\theta\theta}}\frac{\partial P_g}{\partial \theta}
+u^\theta\left(
		u^t \frac{\partial P_g}{\partial t}
		+u^i \frac{\partial P_g}{\partial x^i}
	\right)
\nonumber\\
&&
= -\alpha^2 u^t u^\theta G^t
	+\gamma_{rr}u^r u^\theta G^r
	+\left[1+\gamma_{\theta\theta}(u^\theta)^2\right] G^\theta
	+\gamma_{\phi\phi}(\Omega+\beta^\phi)(G^\phi+\beta^\phi G^t)u^tu^\theta. 
\label{eq:thetaEuler}
\end{eqnarray}
%

In the same way, 
the Euler equation in $\phi$-direction, i.e. momentum conservation 
in $\phi$-direction is calculated as 
%
\begin{eqnarray}
&&
 \rho_0 h_{\rm g} u^t \frac{\partial u^\phi}{\partial t}
+\rho_0 h_{\rm g} u^i \frac{\partial u^\phi}{\partial x^i}
+\rho_0 h_{\rm g} u^t u^r 
	\left\{
		(\Omega+\beta^\phi)\left[
			\left(\ln\gamma_{\phi\phi}\right)_{,r}
			+\frac{\beta_\phi}{\alpha^2}\beta^\phi_{,r}
		\right]
		-2\beta^\phi \left(\ln\alpha\right)_{,r}
		+\beta^\phi_{,r}
	\right\}
\nonumber\\
&&
+\rho_0 h_{\rm g} u^t u^\theta 
	\left\{
		(\Omega+\beta^\phi)\left[
			\left(\ln\gamma_{\phi\phi}\right)_{,\theta}
			+\frac{\beta_\phi}{\alpha^2}\beta^\phi_{,\theta}
		\right]
		-2\beta^\phi \left(\ln\alpha\right)_{,\theta}
		+\beta^\phi_{,\theta}
	\right\}
+\frac{1}{\gamma_{\phi\phi}}\frac{\partial P_g}{\partial \phi}
+\frac{\beta^\phi}{\alpha^2}
	\left(
		\frac{\partial P_g}{\partial t}
		-\beta^\phi \frac{\partial P_g}{\partial \phi}
	\right)
\nonumber\\
&&
+u^\phi\left(
		u^t \frac{\partial P_g}{\partial t}
		+u^i \frac{\partial P_g}{\partial x^i}
	\right)
= -\alpha^2 u^t u^\phi G^t
	+\gamma_{rr}u^r u^\phi G^r
	+\gamma_{\theta\theta}u^\theta u^\phi G^\theta
	+G^\phi
	+\gamma_{\phi\phi}(\Omega+\beta^\phi)(G^\phi+\beta^\phi G^t)u^tu^\phi. 
\label{eq:phiEuler}
\end{eqnarray}
%
%
The for the local energy conservation is obtained from 
$u_\alpha T^{\alpha\beta}_{~~;\beta}=u_\alpha G^\alpha$ calculated as 
%
\begin{eqnarray}
-n u^t \frac{\partial}{\partial t}\left(\frac{\rho_0 h_{\rm g}}{n}\right)
-n u^i \frac{\partial}{\partial x^i}\left(\frac{\rho_0 h_{\rm g}}{n}\right)
+u^t \frac{\partial P_g}{\partial t}
+u^i \frac{\partial P_g}{\partial x^i}
=
-\alpha^2 u^t G^t
+\gamma_{rr}u^r G^r
+\gamma_{\theta\theta}u^\theta G^\theta
+\gamma_{\phi\phi}(\Omega+\beta^\phi)(G^\phi+\beta^\phi G^t)u^t. 
\label{eq:energyEq}
\end{eqnarray}
%
The right hand side of Eq. (\ref{eq:energyEq}) is also calculated 
by the heating and cooling function defined in the comoving frame 
as 
%
\begin{equation}
u_\alpha G^\alpha=-G^{\bar{t}}=\bar{\Lambda}-\bar{\Gamma}. 
\end{equation}
%

\subsubsection{Radiation moment equations}
The radiation moment equation $R^{\alpha\beta}_{~~;\beta}=-G^\alpha$ gives 
the equation for the energy density, the radiation flux and 
the radiation pressure tensor. 
The radiation energy equation is obtained from $t$-component of 
this equation $R^{t\alpha}_{~~;\alpha}=-G^t$ calculated as 
%
\begin{eqnarray}
&&
\frac{\partial R^{tt}}{\partial t}
+\frac{1}{\alpha \sqrt{\gamma}}
	\left[
		\frac{\partial}{\partial r}
			\left(\alpha \sqrt{\mathstrut \gamma}~R^{tr}\right)
		+\frac{\partial}{\partial \theta}
			\left(\alpha \sqrt{\mathstrut \gamma}~R^{t\theta}\right)
	\right]
+\frac{\partial R^{t\phi}}{\partial \phi}
+\left(g^{tt}g_{tt,r}+g^{t\phi}g_{t\phi,r}\right)R^{tr}
+\left(g^{tt}g_{t\phi,r}+g^{t\phi}g_{\phi\phi,r}\right)R^{r\phi}
\nonumber\\
&&
+\left(g^{tt}g_{tt,\theta}+g^{t\phi}g_{t\phi,\theta}\right)R^{t\theta}
+\left(g^{tt}g_{t\phi,\theta}
		+g^{t\phi}g_{\phi\phi,\theta}\right)R^{\theta\phi}
=-G^t. \label{eq:RME0}
\end{eqnarray}
%

%
The radiation momentum equation in $r$-direction 
$R^{r\alpha}_{~~;\alpha}=-G^r$ is calculated as 
%
\begin{eqnarray}
&&
\frac{\partial R^{tr}}{\partial t}
+\frac{1}{\alpha \sqrt{\gamma}}
	\left[
		\frac{\partial}{\partial r}
			\left(\alpha \sqrt{\mathstrut \gamma}~R^{rr}\right)
		+\frac{\partial}{\partial \theta}
			\left(\alpha \sqrt{\mathstrut \gamma}~R^{r\theta}\right)
	\right]
+\frac{\partial R^{r\phi}}{\partial \phi}
+\frac{1}{2}g^{rr}
	\left(
	g_{rr,r}R^{rr}+2g_{rr,\theta}R^{r\theta}
	-g_{\theta\theta,r}R^{\theta\theta}
	\right)
\nonumber\\
&&
-\frac{1}{2}g^{rr}
	\left(
	g_{tt,r}R^{tt}
	+2g_{t\phi,r}R^{t\phi}
	+g_{\phi\phi,r}R^{\phi\phi}
	\right)
=-G^r. \label{eq:RME1}
\end{eqnarray}
%
In the similar manner, 
the radiation momentum equation in $\theta$-direction 
$R^{\theta\alpha}_{~~;\alpha}=-G^\theta$ is calculated as 
%
\begin{eqnarray}
&&
\frac{\partial R^{t\theta}}{\partial t}
+\frac{1}{\alpha \sqrt{\gamma}}
	\left[
		\frac{\partial}{\partial r}
			\left(\alpha \sqrt{\mathstrut \gamma}~R^{r\theta}\right)
		+\frac{\partial}{\partial \theta}
			\left(\alpha \sqrt{\mathstrut \gamma}~R^{\theta\theta}\right)
	\right]
+\frac{\partial R^{\theta\phi}}{\partial \phi}
+\frac{1}{2}g^{\theta\theta}
	\left(
	g_{\theta\theta,\theta}R^{\theta\theta}
	+2g_{\theta\theta,r}R^{r\theta}
	-g_{rr,\theta}R^{rr}
	\right)
\nonumber\\
&&
-\frac{1}{2}g^{\theta\theta}
	\left(
	g_{tt,\theta}R^{tt}
	+2g_{t\phi,\theta}R^{t\phi}
	+g_{\phi\phi,\theta}R^{\phi\phi}
	\right)
=-G^\theta. \label{eq:RME2}
\end{eqnarray}
%
The radiation momentum equation in $\phi$-direction 
$R^{\phi\alpha}_{~~;\alpha}=-G^\phi$ is calculated as 
%
\begin{eqnarray}
&&
\frac{\partial R^{t\phi}}{\partial t}
+\frac{1}{\alpha \sqrt{\gamma}}
	\left[
		\frac{\partial}{\partial r}
			\left(\alpha \sqrt{\mathstrut \gamma}~R^{r\phi}\right)
		+\frac{\partial}{\partial \theta}
			\left(\alpha \sqrt{\mathstrut \gamma}~R^{\theta\phi}\right)
	\right]
+\frac{\partial R^{\phi\phi}}{\partial \phi}
+\left(g^{\phi\phi}g_{t\phi,r}+g^{t\phi}g_{tt,r}\right)R^{tr}
+\left(g^{\phi\phi}g_{\phi\phi,r}+g^{t\phi}g_{t\phi,r}\right)R^{r\phi}
\nonumber\\
&&
+\left(g^{\phi\phi}g_{t\phi,\theta}
		+g^{t\phi}g_{tt,\theta}\right)R^{t\theta}
+\left(g^{\phi\phi}g_{\phi\phi,\theta}
		+g^{t\phi}g_{t\phi,\theta}\right)R^{\theta\phi}
=-G^\phi. \label{eq:RME3}
\end{eqnarray}
%
Finally, 
by inserting the components of the radiation stress tensor given by 
Eq. (\ref{eq:RmomCFbyLNRF}) in to Eqs. (\ref{eq:RME0}), (\ref{eq:RME1}), 
(\ref{eq:RME2}) and (\ref{eq:RME3}), we obtain the radiation moment 
equations corresponding to Eqs. (49), (52), (53) and (54) in \cite{P06} as 
%
\begin{eqnarray}
&&
\frac{\partial}{\partial t}
	\left( 
		\frac{\hat{E}}{\alpha^2} 
	\right)
+\frac{1}{\alpha \sqrt{\mathstrut \gamma}}
		\frac{\partial}{\partial r}
			\left( 
				\sqrt{\frac{\gamma}{\gamma_{rr}}}\hat{F}^r	
			\right)
+\frac{1}{\alpha \sqrt{\mathstrut \gamma}}
		\frac{\partial}{\partial \theta}
			\left(
				\sqrt{\frac{\gamma}{\gamma_{\theta\theta}}}
				\hat{F}^\theta		
			\right)
+\frac{\partial}{\partial \phi}
	\left[
			\frac{1}{\alpha}
			\left(
				\frac{\hat{F}^\phi}{\sqrt{\mathstrut \gamma_{\phi\phi}}}
				-\beta^\phi
					\frac{\hat{E}}{\alpha}
			\right)						
	\right]
\nonumber\\
&&
+2\left(\ln\alpha\right)_{,r}
	\frac{\hat{F}^r}{\alpha\sqrt{\mathstrut \gamma_{rr}}} 
+2\left(\ln\alpha\right)_{,\theta}
	\frac{\hat{F}^\theta}{\alpha\sqrt{\mathstrut \gamma_{\theta\theta}}}
-\frac{\beta^\phi_{,r}}{\alpha^2}
	\sqrt{\frac{\gamma_{\phi\phi}}{\gamma_{rr}}}\hat{P}^{r\phi}
-\frac{\beta^\phi_{,\theta}}{\alpha^2}
	\sqrt{\frac{\gamma_{\phi\phi}}{\gamma_{\theta\theta}}}
	\hat{P}^{\theta\phi}
=-G^t. \label{eq:RME0P}
\\
&&
\frac{\partial }{\partial t}
	\left( 
		\frac{\hat{F}^r}{\alpha\sqrt{\mathstrut \gamma_{rr}}}
	\right)
+\frac{1}{\alpha \sqrt{\mathstrut \gamma}}
		\frac{\partial}{\partial r}
			\left(\alpha \sqrt{\mathstrut \gamma} 
				\frac{\hat{P}^{rr}}{\gamma_{rr}}
			\right)
+\frac{1}{\alpha \sqrt{\mathstrut \gamma}}
		\frac{\partial}{\partial \theta}
			\left(\alpha \sqrt{\mathstrut \gamma} 
				\frac{\hat{P}^{r\theta}}
					{\sqrt{\mathstrut \gamma_{rr}\gamma_{\theta\theta}}}
			\right)
+\frac{\partial }{\partial \phi}
	\left[ 
			\frac{1}{\sqrt{\mathstrut \gamma_{rr}}}
			\left(
				\frac{\hat{P}^{r\phi}}{\sqrt{\mathstrut \gamma_{\phi\phi}}}
				-\beta^\phi
					\frac{\hat{F}^r}{\alpha}
			\right)
	\right]
\nonumber\\
&&
+\frac{\left(\ln\alpha\right)_{,r}}{\gamma_{rr}}\hat{E}
-\frac{\sqrt{\mathstrut\gamma_{\phi\phi}}\beta^\phi_{,r}}
	{\alpha\gamma_{rr}}\hat{F}^\phi
+\frac{1}{2}\left(\ln\gamma_{rr}\right)_{,r} 
			\frac{\hat{P}^{rr}}{\gamma_{rr}}
+\left(\ln\gamma_{rr}\right)_{,\theta} 
			\frac{\hat{P}^{r\theta}}
				{\sqrt{\mathstrut \gamma_{rr}\gamma_{\theta\theta}}}	
-\frac{\left(\ln\gamma_{\theta\theta}\right)_{,r}}{2\gamma_{rr}}
		\hat{P}^{\theta\theta} 
-\frac{\left(\ln\gamma_{\phi\phi}\right)_{,r}}{2\gamma_{rr}}
		\hat{P}^{\phi\phi} 
=-G^r. \label{eq:RME1P}
\\
&&
\frac{\partial}{\partial t}
	\left( 
			\frac{\hat{F}^\theta}
					{\alpha\sqrt{\mathstrut \gamma_{\theta\theta}}}	
	\right)
+\frac{1}{\alpha \sqrt{\mathstrut \gamma}}
		\frac{\partial}{\partial r}
			\left(\alpha \sqrt{\mathstrut \gamma} 
			\frac{\hat{P}^{r\theta}}
				{\sqrt{\mathstrut \gamma_{rr}
						\gamma_{\theta\theta}}}				
			\right)
+\frac{1}{\alpha \sqrt{\mathstrut \gamma}}
		\frac{\partial}{\partial \theta}
			\left(\alpha \sqrt{\mathstrut \gamma} 
			\frac{\hat{P}^{\theta\theta}}{\gamma_{\theta\theta}}		
			\right)
+\frac{\partial}{\partial \phi}
		\left[ 
			\frac{1}{\sqrt{\mathstrut\gamma_{\theta\theta}}}
			\left(
				\frac{\hat{P}^{\theta\phi}}
					{\sqrt{\mathstrut \gamma_{\phi\phi}}}
				-\beta^\phi\frac{\hat{F}^\theta}{\alpha}
			\right)			
		\right]
\nonumber\\
&&
+\frac{\left(\ln\alpha\right)_{,\theta}}{\gamma_{\theta\theta}}\hat{E}
-\frac{\sqrt{\mathstrut \gamma_{\phi\phi}}\beta^\phi_{,\theta}}
		{\alpha\gamma_{\theta\theta}}
	\hat{F}^\phi
-\frac{1}{2}\left(\ln\gamma_{rr}\right)_{,\theta}
	\frac{\hat{P}^{rr}}{\gamma_{\theta\theta}}
+\left(\ln\gamma_{\theta\theta}\right)_{,r}
			\frac{\hat{P}^{r\theta}}
				{\sqrt{\mathstrut \gamma_{rr}\gamma_{\theta\theta}}}	
+\frac{1}{2}\left(\ln\gamma_{\theta\theta}\right)_{,\theta}
	\frac{\hat{P}^{\theta\theta}}{\gamma_{\theta\theta}}
-\frac{1}{2}\left(\ln\gamma_{\phi\phi}\right)_{,\theta}
	\frac{\hat{P}^{\phi\phi}}{\gamma_{\theta\theta}}
=-G^\theta. \label{eq:RME2P}
\\
&&
\frac{\partial}{\partial t}
	\left[ 
			\frac{1}{\alpha}
			\left(
				\frac{\hat{F}^\phi}{\sqrt{\mathstrut \gamma_{\phi\phi}}}
				-\beta^\phi
					\frac{\hat{E}}{\alpha}
			\right)	
	\right]
+\frac{1}{\alpha \sqrt{\gamma}}
		\frac{\partial}{\partial r}
			\left[ 
				\frac{\alpha \sqrt{\mathstrut \gamma}}
					{\sqrt{\mathstrut \gamma_{rr}}}
				\left(
					\frac{\hat{P}^{r\phi}}
							{\sqrt{\mathstrut \gamma_{\phi\phi}}}
					-\beta^\phi
						\frac{\hat{F}^r}{\alpha}
				\right)
			\right]
\nonumber\\
&&
+\frac{1}{\alpha \sqrt{\gamma}}
		\frac{\partial}{\partial \theta}
			\left[ 
				\frac{\alpha \sqrt{\mathstrut \gamma}}
						{\sqrt{\mathstrut \gamma_{\theta\theta}}}
				\left(
					\frac{\hat{P}^{\theta\phi}}
							{\sqrt{\mathstrut \gamma_{\phi\phi}}}
					-\beta^\phi\frac{\hat{F}^\theta}{\alpha}
				\right)
			\right]
+\frac{\partial}{\partial \phi}
	\left[ 
		\frac{\hat{P}^{\phi\phi}}{\gamma_{\phi\phi}}
			-\frac{2\beta^\phi}{\alpha\sqrt{\mathstrut \gamma_{\phi\phi}}}
				\hat{F}^\phi
			+\left(\frac{\beta^\phi}{\alpha}\right)^2 \hat{E}
	\right]
\nonumber\\
&&
+\left[\beta^\phi_{,r}-2\beta^\phi(\ln\alpha)_{,r}\right]
			\frac{\hat{F}^r}{\alpha\sqrt{\mathstrut \gamma_{rr}}} 
+\left[\beta^\phi_{,\theta}-2\beta^\phi(\ln\alpha)_{,\theta}\right]
			\frac{\hat{F}^\theta}
					{\alpha\sqrt{\mathstrut \gamma_{\theta\theta}}}
+\left[
	\left(\ln \gamma_{\phi\phi}\right)_{,r}
	+\frac{\beta_\phi \beta^\phi_{,r}}{\alpha^2}
\right]
	\frac{\hat{P}^{r\phi}}{\sqrt{\mathstrut \gamma_{rr}\gamma_{\phi\phi}}}
\nonumber\\
&&
+\left[
	\left(\ln \gamma_{\phi\phi}\right)_{,\theta}
	+\frac{\beta_\phi \beta^\phi_{,\theta}}{\alpha^2}
\right]
	\frac{\hat{P}^{\theta\phi}}
			{\sqrt{\mathstrut \gamma_{\theta\theta}\gamma_{\phi\phi}}}
=-G^\phi, \label{eq:RME3P}
\end{eqnarray}
%
where the derivatives for $\ln \alpha$, $\beta^\phi$, $\ln \gamma_{rr}$, 
$\ln \gamma_{\theta\theta}$ and $\ln \gamma_{\phi\phi}$ 
with respect to $r$ and $\theta$ are given in App. \ref{app:metric}.

%

%

%

\section{Concluding remarks}
As is well known, the moment equations which are truncated at the 
finite order do not constitute a complete system of equations. 
That is, the number of equations are smaller than the number of variables 
to be solved. 
Thus, we need the additional equations to close the system of 
the equations \citep{C60,M70,P73,RL79,MM84,S91} 
such as the Eddington factors.  
In the diffusion limit where the photon mean-free path is much smaller 
than the characteristic length of the system, i.e. in the optically 
thick region, the so-called Eddington approximation 
$P^{ij}=(\delta^{ij}/3)E$ is valid. 
On the other hand, in the optically thin region, 
the radiation becomes anisotropic in general. 
So, in such case, the full angle-dependent radiative transfer equations 
should be solved, and sometimes  
the variable Eddington factors are introduced. 
Some authors have expressed the variable Eddington factors 
as a function of the optical depth. 
Additionally, it is pointed out that the influence of the flow velocity 
should be also taken into account when constructing the closure relations. 
These problems and their calculation methods are extensively studied 
in the past studies 
\citep{AM70,HR71,TTKO75,M80,MM84,SB89,NTZ93,YM95,KFM98,F06}. 
The similar analysis for the closure relations will be studied 
in the case of the Kerr space-time in future. 
In the present study, we use the Boyer-Lindquist coordinate which is a 
frequently used coordinate in the astrophysics around a rotating black hole. 
However, there is a coordinate singularity at the horizon 
in this coordinate. 
Future studies also resolve this singularity by using the another coordinate 
with no singularity at the horizon (see, e.g. Takahashi 2007). 
In this study, we have derived the radiation hydrodynamic equation in the 
Kerr space-time. 
While the interactions between the matter and the radiation are defined and 
calculated in the comoving frame, the derivatives which are 
used to describe the global evolutions of both the matter and the radiation 
are calculated in the BLF.  
Both the matter and the radiation are influenced by the frame-dragging 
due to the black hole's rotation. 
In this approach, as a fixed orthonormal frame, we use 
the so-called locally non-rotating reference frame (LNRF) which is 
firstly calculated in the Kerr space-time by \cite{BPT72} by using 
the Boyer-Lindquist coordinate. 
The special relativistic effects such as beaming effects are introduced 
when the Lorentz transformation between the LNRF and the comoving frame. 
On the other hand, the general relativistic effects such as 
frame-dragging effects due the black hole's rotation and the gravitational 
redshift are introduced by the tetrads describing 
the BLF.  
%

\section*{Acknowledgments}
The author is grateful to 
Professors Y. Eriguchi and S. Mineshige for their continuous encouragements,
and J. Fukue, Y. Sekiguchi, M. Shibata for useful discussion and comments, 
and J. Thiel for proofreading. 
The author also thanks to Department of Physics at Montana State University 
for its hospitality and Sachiko Tsuruta for her hospitality. 
This research was partially supported by the Ministry of Education,
Culture, Sports, Science and Technology, Grant-in-Aid for 
Japan Society for the Promotion of Science (JSPS) Fellows (17010519).


\appendix

\section{Metric Components and Differential Values}
\label{app:metric}
Here, we present the explicit expressions for the metric components and 
the their differential values  
with respect to $r$ and $\theta$ used in this paper. 
Non-zero components of the metric are given as  
%
\begin{eqnarray}
&&
g_{tt}=-\alpha^2+\beta_\phi \beta^\phi=-\left(1-\frac{2mr}{\Sigma}\right),~~~
g_{t\phi}=\beta_\phi=-\frac{2mar\sin^2\theta}{\Sigma},~~~
g_{\phi\phi}=\gamma_{\phi\phi}=\frac{A\sin^2\theta}{\Sigma},\nonumber\\
&&
g_{rr}=\gamma_{rr}=\frac{\Sigma}{\Delta},~~~
g_{\theta\theta}=\gamma_{\theta\theta}=\Sigma,
\end{eqnarray}
%
and
%
\begin{eqnarray}
&&
g^{tt}=-\frac{1}{\alpha^2}=-\frac{A}{\Sigma\Delta},~~~
g^{t\phi}=\frac{\beta^\phi}{\alpha^2}=-\frac{2mar}{\Sigma\Delta},~~~
g^{\phi\phi}=\gamma^{\phi\phi}-\frac{(\beta^\phi)^2}{\alpha^2}
	=\frac{1}{\Delta\sin^2\theta}\left(1-\frac{2mr}{\Sigma}\right),
\nonumber\\
&&
g^{rr}=\gamma^{rr}=\frac{\Delta}{\Sigma},~~~
g^{\theta\theta}=\gamma^{\theta\theta}=\frac{1}{\Sigma}. 
\end{eqnarray}
%
The differential values of the metric used in this paper are given as
%
\begin{eqnarray}
&&
\left(\ln \alpha \right)_{,r}=
\frac{1}{2}\left[
	\left(\ln \Sigma \right)_{,r}
	+\left(\ln \Delta \right)_{,r}
	-\left(\ln A \right)_{,r}
\right],~~~
\beta^\phi_{,r}=-\omega\left[
			\frac{1}{r}-\left(\ln A\right)_{,r}
		\right],~~~
\nonumber\\
&&
\left(\ln \gamma_{rr} \right)_{,r}=
	\left(\ln \Sigma \right)_{,r}+\left(\ln \Delta \right)_{,r},~~~
\left(\ln \gamma_{\theta\theta} \right)_{,r}=
	\left(\ln \Sigma \right)_{,r},~~~
\left(\ln \gamma_{\phi\phi} \right)_{,r}=
	\left(\ln A\right)_{,r}
	-\left(\ln \Sigma \right)_{,r},
\nonumber\\
&&
\left(\ln \alpha \right)_{,\theta}=
\frac{1}{2}\left[
	\left(\ln \Sigma \right)_{,\theta}
	+\left(\ln \Delta \right)_{,\theta}
	-\left(\ln A \right)_{,\theta}
\right],~~~
\beta^\phi_{,\theta}=\omega\left(\ln A\right)_{,\theta},~~~
\nonumber\\
&&
\left(\ln \gamma_{rr} \right)_{,\theta}=
	\left(\ln \Sigma \right)_{,\theta}
	+\left(\ln \Delta \right)_{,\theta},~~~
\left(\ln \gamma_{\theta\theta} \right)_{,\theta}=
	\left(\ln \Sigma \right)_{,\theta},~~~
\left(\ln \gamma_{\phi\phi} \right)_{,\theta}=
	\left(\ln A\right)_{,\theta}+\frac{2}{\tan\theta}
	-\left(\ln \Sigma \right)_{,\theta},~~~ 
\end{eqnarray}
%
where 
%
\begin{eqnarray}
&&
\left(\ln \Sigma \right)_{,r}=\frac{2r}{\Sigma},~~~
\left(\ln \Delta \right)_{,r}=\frac{2(r-m)}{\Delta},~~~
\left(\ln A \right)_{,r}=
	\frac{2}{A}\left[(r-m)\Sigma+(r+m)(r^2+a^2)\right],
\nonumber\\
&&
\left(\ln \Sigma \right)_{,\theta}=-a^2\sin 2\theta,~~~
\left(\ln \Delta \right)_{,\theta}=0,~~~
\left(\ln A \right)_{,\theta}=-\frac{a^2\Delta}{A}\sin 2\theta. 
\end{eqnarray}
%
Here, we also give the differential values of the metric as
%
\begin{eqnarray}
&&g_{tt,r}=\frac{2m}{\Sigma}\left(1-\frac{2r^2}{\Sigma}\right),~~~
g_{t\phi,r}=\frac{2ma\sin^2\theta}{\Sigma}
	\left(\frac{2r^2}{\Sigma}-1\right),~~~
g_{\phi\phi,r}=2\sin^2\theta\left[
		r-m+\frac{(3r^2+a^2)m}{\Sigma}-\frac{2mr^2(r^2+a^2)}{\Sigma^2}
	\right],\nonumber\\
&&
g_{rr,r}=\frac{2}{\Delta}\left[r-\frac{(r-m)\Sigma}{\Delta}\right],~~~
g_{\theta\theta,r}=2r,
\end{eqnarray}
%
and 
%
\begin{eqnarray}
&&
g_{tt,\theta}=\frac{2ma^2r}{\Sigma^2}\sin 2\theta,~~~
g_{t\phi,\theta}=-\frac{2mar(r^2+a^2)}{\Sigma^2}\sin 2\theta,~~~
g_{\phi\phi,\theta}=\left[
		\Delta+\frac{2mr(r^2+a^2)^2}{\Sigma^2}
	\right]\sin 2\theta,\nonumber\\
&&
g_{rr,\theta}=-\frac{a^2}{\Delta}\sin 2\theta,~~~
g_{\theta\theta,\theta}=-a^2\sin 2\theta. 
\end{eqnarray}
%


\bsp

\label{lastpage}

\end{document}